\documentclass
{raa}            

\usepackage{graphicx,times}             

\begin{document}

   \title{Investigation of radio pulsar emission features using power spectra}
   \volnopage{Vol.0 (20xx) No.0, 000--000}      
   \setcounter{page}{1}          

   \author{V.M.Malofeev
      \inst{1}
   \and S.A.Tyul'bashev
      \inst{1}
   }

   \institute{P.N.Lebedev Physical institute of the Russian Academy of Sciences, V.V.Vitkevich Pushchino radioastronomy observatory,\\ PRAO ASC FIAN, Pushchino, Russia}

   \date{Received~~20 month day; accepted~~20~~month day}

\abstract{Since 2013 round-the-clock monitoring of the sky is carried out simultaneously in 96 space beams using
the high-sensitivity radio telescope of LPA (Large Phased Array) at the frequency 110.25 MHz. These observations
are made under the program of the interplanetary plasma investigation. The same data is used for the  search for
pulsars by means of the power spectra. For the increase of the pulsar search sensitivity the summation to 500-600
of power spectra corresponding to the different days of the observations is made. In the integrated spectra of
the known pulsars, besides expected improvement of a signal-to-noise ($S/N$) ratio for the frequency harmonics,
were showed some features which are explored in the paper. We present 27 strongest pulsars which are in the field
of declinations $21\degr - 42\degr$ at which the connection of observable details in the integrated power spectra
with the presence of pulsar periods of the second ($P_2$) and third ($P_3$) class has been discovered.
The empirical relations for the calculation of these periods are obtained. The value $P_2$ is estimated  for 26
pulsars, and  for 15 sources it is made for the first time. The value $P_3$ is estimated for 13 pulsars, from
them for 5 sources they are given for the first time.
}

   \authorrunning{V.M.Malofeev \& S.A.Tyul'bashev}
   \titlerunning{Investigation of radio pulsar emission features using power spectra}
   \maketitle

%

\section{Introduction}

Fourier  power spectrum is a good tool for searching for frequency of the periodical processes, therefore it is no
wonder that search for pulsars is made, as a rule, with the use of power spectra. The Fourier transform is made or
for the correlation function, or the Fourier transform of a signal is raised to a square. After detection of the
harmonics in a spectrum the integration of a signal is made with a period corresponding to the inverse ratio of the
first harmonic frequency. Thus search for a dispersion measure is carried out, in a case of observations in the
several frequency channels. As a result of averaging the mean profile of a pulsar is obtained, which is the
narrowest for the "correct" dispersion measure. The recurring observations using the same telescope by which the
candidates for pulsars are is found, in addition to other frequency, are even better than observations by other
telescopes, and it is the reliable method to confirm the pulsar existence. With small variations such organization of
operation on the pulsar search is accepted everywhere, see the last review of Barr et~al.~(\cite{B13}). The power spectrum as the
such method of the search for pulsars is only auxiliary means, and in the further operation on the pulsar
investigation is not used, as a rule.

In 2013 there is begun a round-the-clock monitoring of the sky by the radio telescope LPA under the program of
investigation of  an interplanetary plasma, and especially coronal mass ejections (Shishov et~al.~\cite{Sh16}); project
"Space Weather"). Besides of this program the review data was used for the search for pulsars also (Tyul'bashev et~al.~\cite{T16}, ~\cite{T17a});
project "BSA-Analytics", http://bsa-analytics.prao.ru ). As pulsars are the objects with the small flux densities
by their search the main problem is a realization of the highest possible sensitivity of observations so that it
was possible to find out extremely weak objects. Taking into account the daily monitoring as an obvious method of
increasing of the sensitivity, the accumulation of the power spectra is made, corresponding to the coordinates of
the same points in the sky. In such integrated spectra, expected improvement of a $S/N$  ratio for the frequency
harmonics and the appearance more their numbers (up to 108 for PSR J0323+3944), the appearance of the several
features demanding an explanation was unexpected. In this paper the analysis of the integrated spectra is carried
out and the nature of two observable features of power spectra is considered.

\section{Observations and data processing}

Observations were made using the Large Phased Array of the P.N.Lebedev Physical Institute at the centre frequency
110.25 MHz. LPA is the phased array constructed on the half-wave dipoles, the geometrical area is 72000 m$^2$ and
the effective area is about 45000 m$^2$ in a direction to zenith in the configuration used by us. The size of one
beam is approximately $0.5\degr \times 1\degr$. The antenna is the meridian instrument, therefore any direction can be
observed only once a day on an extent approximately 7 minutes. The details, concerning the radio telescopes
created on the basis of the antenna LPA, are given in the papers Shishov et~al.~(\cite{Sh16}), Tyul'bashev et~al.~(\cite{T16}). There are two basic modes of a
data recording. The first mode is a record in the six frequency channels with the channel bandwidth of 430 kHz and
the data-sampling interval  0.1 s. The second mode is a record in a 32-channel frequency mode with the bandwidth
of the channel 78 kHz and the data-sampling interval 12.5 ms. Since August, 2014 both modes are used
simultaneously during observations in 96 beams of the antenna, overlapping the declinations
$-9\degr < \delta < 42\degr $. Data presented in this paper were obtained on the declinations
$21\degr < \delta < 42\degr $, where we have the peak sensitivity of the radio telescope and the minimum of
noises is observed.

Before a calculation a power spectrum, cleaning from the pulse interferences was made and a quantity of the noise
level was estimated. If this level was too big, that is more than the threshold level specially calculated for the
given direction, the further operation was not use. In practice it has appeared that approximately from 20 to
$30\%$ of all records are rejected by the primary search for pulsars. The part of this rejected data can be used further
for checkout of the search results and, probably, the pulsar investigations. The power spectra were under
construction both for the 6-channel "small", and for the 32-channel "big" data. To obtain the Fourier power
spectra fast Fourier transform was used. Temporary the file got out in length of 2048 points for "small" and
16384 points for "big" data, what was approximately half of the temporal beam of LPA. Further the spectra obtained
independently for each frequency channel, were sum up and used for a primary search for the periods of new
pulsars (Tyul'bashev et~al.~\cite{T17a}). The following stage was a summation of the power spectra for the different days
corresponding to one direction on celestial sphere. At the moment of a beginning of a processing with "big" data
already two years of the observations was available. There were 500-600 on overage files with the individual spectra to obtain the integrated power spectrum. That was expected 22-24 times increasing of
the $S/N$ ratio for the harmonics of such integrated spectrum.

What it would be possible to expect from such summation? As radiation of the majority of pulsars exteriorly is
represented a set of delta functions and in the obtained power spectrum on an exit there should be a set of delta
functions. In practice a series of observations is limited on time and consequently an average power spectrum the
harmonics should be built on the height, where the first harmonic one is the strongest. In real, such pattern is
broken for an individual power spectra, as pulses of pulsars have different intensity, up to lack of a part of
pulses more often. Frequently unsuccessful subtraction of a base line in the initial data and the strong
low-frequency noise lead to that S/N ratio in the first harmonic is lower, than in the second one.

As a result of summation of an individual power spectra we expected to see a certain "ideal" spectrum. However,
effects of averaging have appeared absolutely unexpected almost for all pulsars of our list. So for 26 from 27
pulsars a modulation of amplitudes of harmonics with the quasi-periods from tens to hundreds milliseconds is
observed. In fig.1 the example of an integrated (a) and individual (b)  power spectrum of the pulsar J0528+2200
in the frequency interval to 40 Hz or 25 ms is given.

\begin{figure}
\includegraphics[width=\columnwidth]{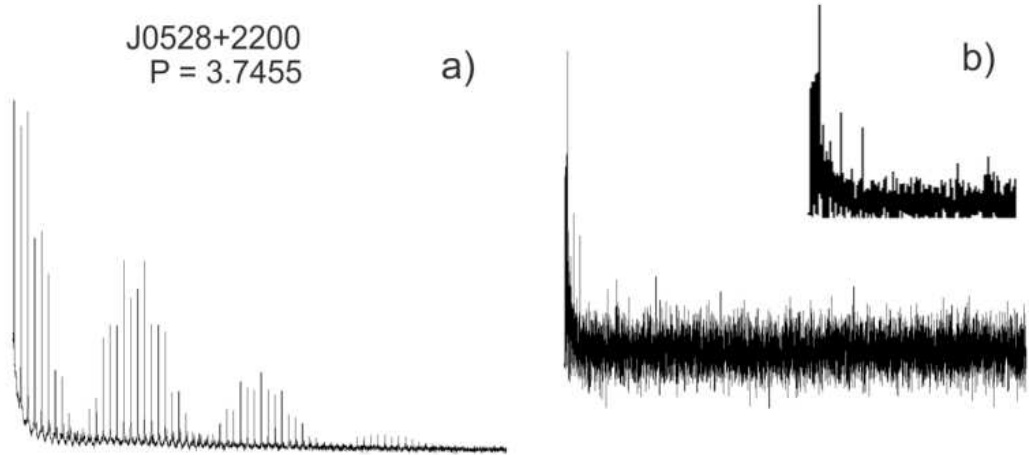}
\caption{Example of an integrated (a) and individual (b)  power spectrum of pulsar J0528+2200. On insert fig.1b
 a fragment of a spectrum for the best representation of the first harmonics is given}
\label{fig:1}
\end{figure}

From fig.1 it is visible that the individual spectrum does not show any features. In an integrated spectrum
a wavy structure is obviously expressed. A modulation showing periodic increasing of a $S/N$ ratio at far
harmonics is presented. The pulsar is constantly visible practically during the most part of days of the
monitoring. Its flux density at 400 MHz according to catalogue ATNF is $S_{400} = 57$~mJy, and at the
frequency of 102 MHz $S_{102} = 100$~mJy (Malofeev  et~al.~\cite{M00}).

By detail consideration of the integrated power spectra, especially the first harmonics, the second feature has
been found out. In fig.2a the integrated power spectrum of pulsar J0323+3944, and also a part of this spectrum in
the large scale, including 10 first harmonics (fig.2b), is presented. The harmonics satellites near to pulsar main harmonic
companions are visible clearly. Earlier these details in the power spectra were not mention by nobody.

\begin{figure}
\includegraphics[width=\columnwidth]{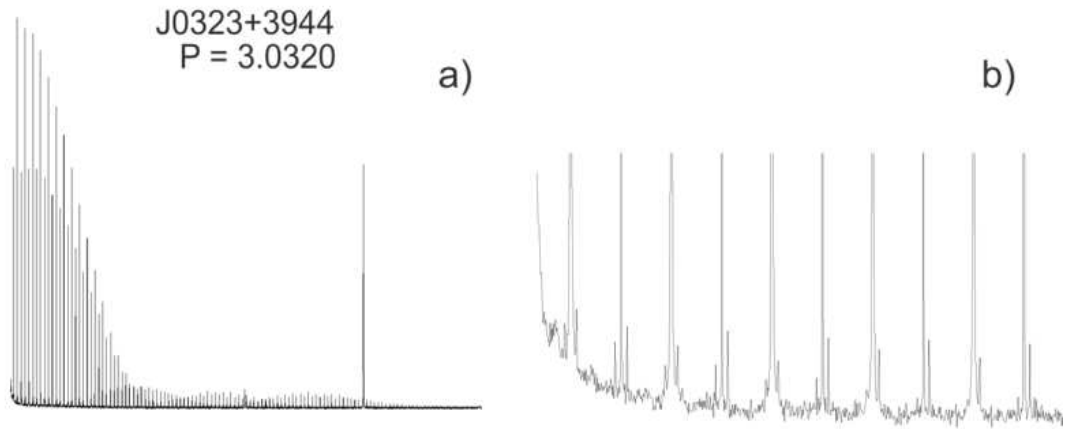}
\caption{The integrated power spectrum of pulsar J0323+3944. On a spectrum regularly appearing interferences of a man made origin (50 and 33 ms) are showed; b) a fragment from a pulsar spectrum - first ten harmonics. The satellites of first harmonics are visible}
\label{fig:2}
\end{figure}

In fig.2a it is visible that the harmonics amplitudes have obviously wavy structure. For the first harmonics there
is an obvious slope of power in harmonics, and the modulation, as in fig.1a, but less expressed is visible.
In fig.2b it is well visible that near to the harmonics, there are the harmonics satellites which are not multiple
to a pulsar period. One of these satellites is to the left, and another to the right of a main harmonic, but
distance to the left and to the right satellites is not identical. It is interesting that both satellites are
visible not in all pulsars, one of satellites is more often visible only. Presence of such details is observed in
13 pulsars from 27 ones.

The power spectra of 25 pulsars are presented in fig.3, and 2 more pulsars are in fig.1, 2. Figures are prepared
so that on an axis of the frequencies all were limited till the presence of the visible harmonics. Therefore the
extreme right frequency on all power spectra is the different. The value of a period given together with the name
of a pulsar, allows be oriented in frequency domain, as the frequency of the first harmonic is equal to the
quantity return to a rotating period of a pulsar. Continuation of fig.3 is fig.4 showing the fragments of the power
spectra of several pulsars with clearly visible harmonics satellites, except of pulsar J0323+3944 presented in
fig.2b.

Not all spectra in fig.3 and fig.4 show an identical collection of the details given in fig.1-2, but all have any
typical features. The analysis of the integrated spectra is made in the following paragraph.

\newpage
\begin{figure*}
\includegraphics[width=16cm]{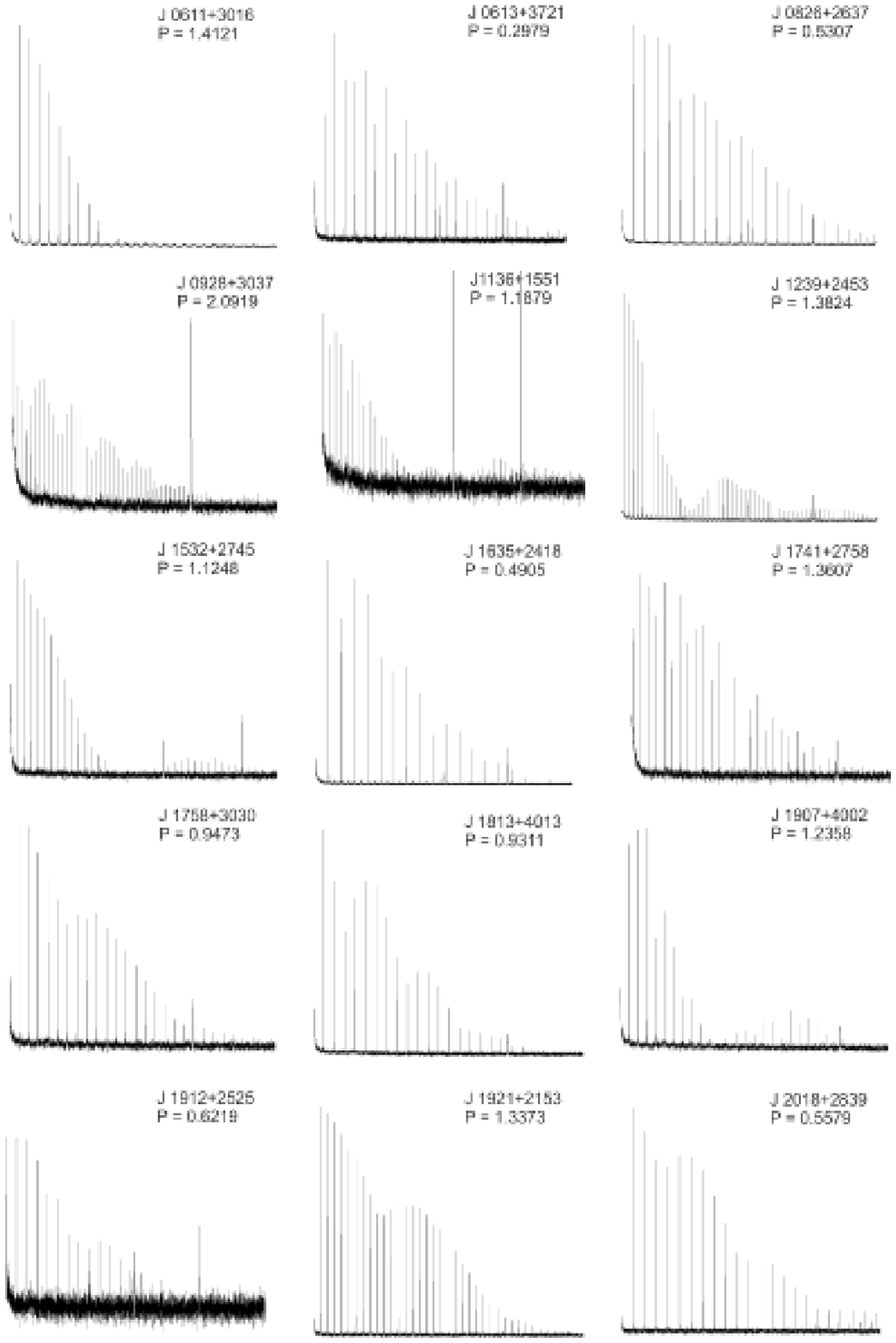}
\end{figure*}
\begin{figure}
\includegraphics[width=\columnwidth]{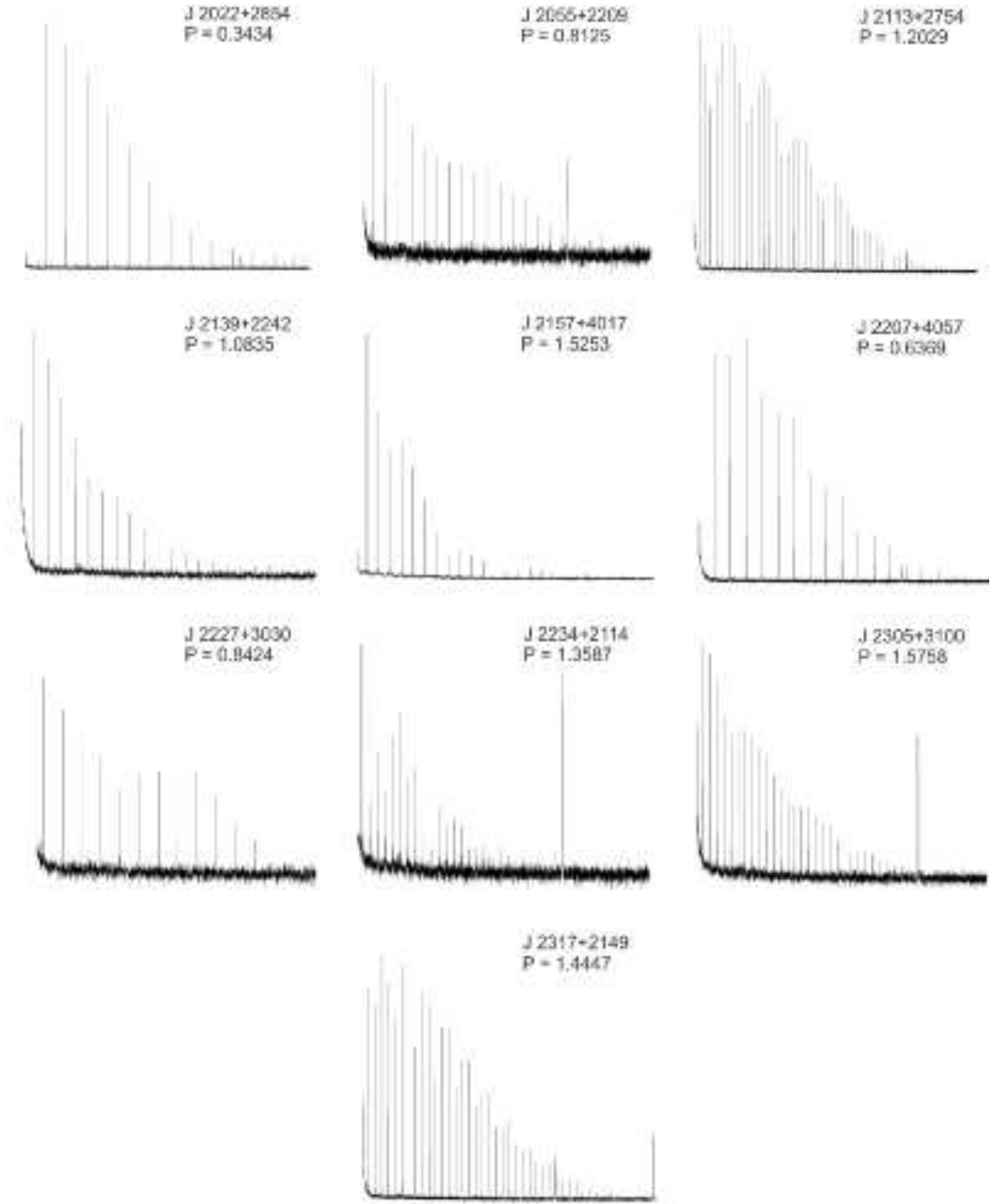}
\caption{Integrated power spectra of 25 pulsars}
\label{fig:1}
\end{figure}

\begin{figure}
\includegraphics[width=\columnwidth]{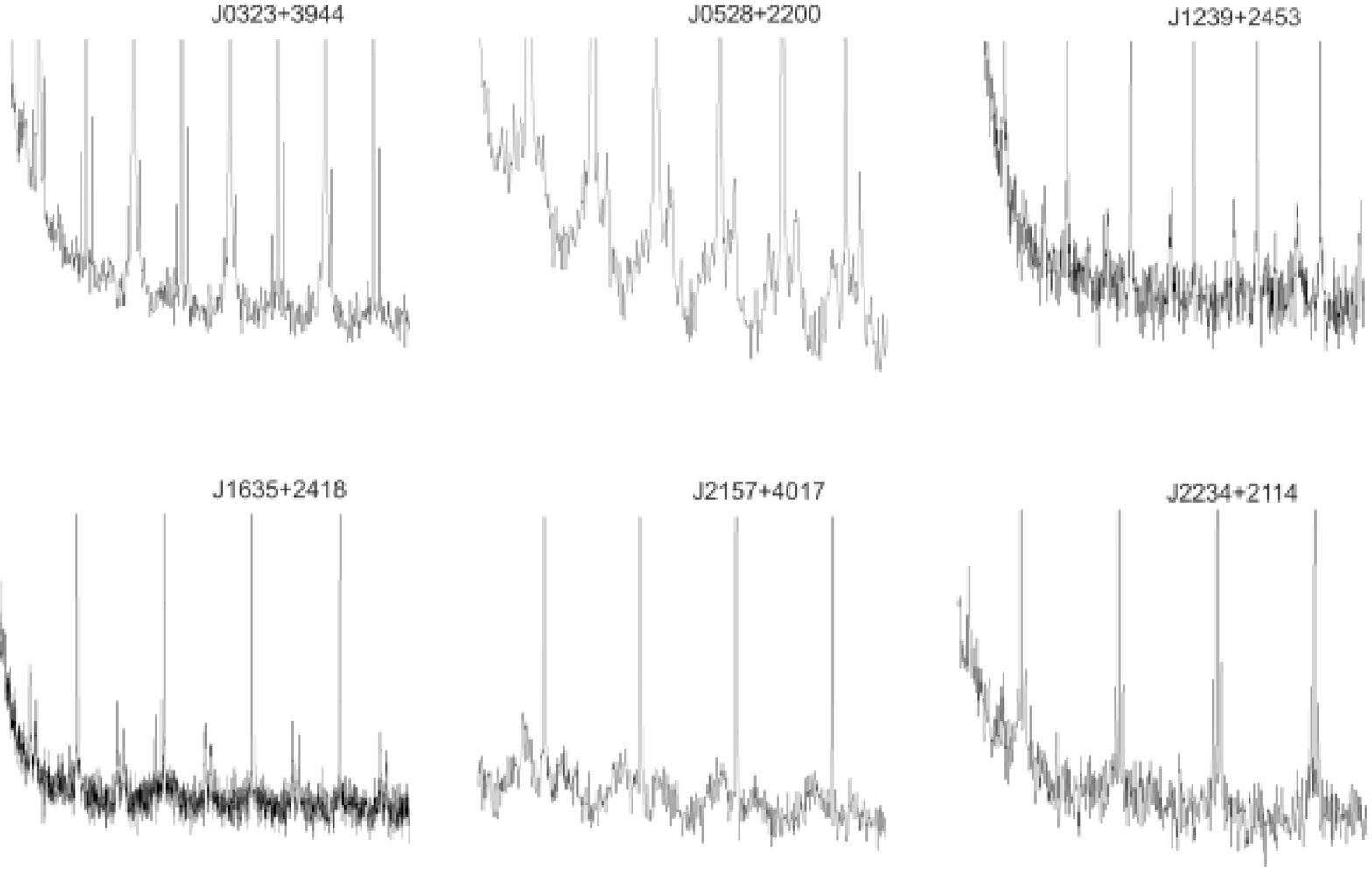}
\caption{The  fragments of the integrated power spectra of pulsars are presented, with the most expressed harmonics satellites, except of pulsar J0528+2200 which is in fig.1b}
\label{fig:1}
\end{figure}

\section{Investigation of power spectra}

As it is known, besides a rotation period ($P_1$) in some of pulsars other periodic or quasi-periodic features
which are not related directly to a main period are observed also. It is, so-called, "the drifting subpulse
phenomenon" which is characterized by two periods $P_2$ and $P_3$. At the very beginning of pulsar investigation
it was discovered (Drake \& Craft~\cite{DC68}) that there were pulsars with inner structure of an individual pulse - subpulses
which showed the regular drift of a phase of arrival inside the so-called "pulsar window", forming "the drift
bands" on the diagram: the number of a period (axis of ordinates) -- a phase of a period (abscissa axis).
The $P_2$  is accepted to term as a period of the second class. The value $P_2$ is the horizontal drift band separation
in time units.  A period of the third class ($P_3$), this distance is determined in number of periods $P_1$ on an
axis of ordinates. Drift to the back end of  a "window" terms as
positive, and to the forward one as negative. One of the first pulsars with the very regular and bright drift
(PSR J0814+7429)  has been found out in Pushchino Vitkevich \& Shitov~(\cite{V70}). Now the drift of subpulses is known in about 70
pulsars and, approximately, for the same number  $P_2$ and $P_3$ are measured (Weltevrede et~al.~\cite{W06}).

There are some models which are attempt to explain the drift effect. The most known: not radial pulsations of a
neutron star Ruderman~(\cite{R68}), the sparking gap model over a polar cap Ruderman \& Sutherland~(\cite{RS75}), rotating of a radiation pattern
around a magnetic axis Sieber \& Oster~(\cite{OS75}), developed in "a rotating carousel" model Deshpande \& Rankin~(\cite{DR99}), and its application
to PSR B0943+10 Rankin et~al.~(\cite{RSD03}), and, at last, a feedback model is proposed by Wright~(\cite{W03}). Unfortunately, not any
of models does not explain all variety of the details related to the effect of subpulse drifts.

We have tried to explain the features showed in fig.1-4 in the integrated power spectra by the presence of the
$P_2$ and the $P_3$. We take, for an example, two pulsars J0528+2200 (fig.1) and J0323+3944 (fig.2), which,
according to the investigation of Weltevrede et~al.~(\cite{W06}) have these periods. This paper, containing the most
measurements of the drifting subpulse phenomenon, was taken as basic for the comparison with our data. It presents
the result of detailed investigations of 187 pulsars at the frequency of 1420 MHz spent in Westerbork (Netherlands).
The one-dimensional and the two-dimensional Fourier power spectra in a pulse window have been used, for measuring
as the modulation coefficient, and $P_2$ and $P_3$. In this analysis the records containing of some thousands of
the pulses were used, as a rule. In fig.1a it is visible that the middle or a maximum of the first "hump" of
modulation in a power spectrum is necessary on a harmonic under the number 19 that corresponds to a period 200 ms.
According to Weltevrede et~al.~(\cite{W06}) in this pulsar a $P_2$ is or -200 ms, or +500 ms. Our estimate allows select the
single-valued of a period 200 ms. We cannot estimate a sign of  the drift. Thus, the period of the
second class is calculated from a simple relation:

\begin{equation}
P_2 = P_1 / n,
\label{eq1}
\end{equation}

\noindent where $n$ it is the harmonic number at which the middle of the first "hump" of modulation in an average
power spectrum takes place, and $P_1$ is a rotation pulsar period. Analyzing the known periods of the third class
of some pulsars, we have obtained the empirical formulas for these satellites of the main harmonic:

\begin{equation}
\begin{array}{lcr}
P_l (n) & = & P_1 (n)  \left( n \times \frac{P_3}{(n \times P_3 - 1)} \right)\\
P_r (n) & = & P_1 (n) \left( n \times \frac{P_3}{(n \times P_3 + 1)} \right),\\
\end{array}
\label{eq2}
\end{equation}
\noindent where $P_l$ is a period of the left satellite, $P_r$ is a period of the right one and $P_3$ is a period
of the third class, in terms of $P_1$.  It is follows from relations (2) that:

\begin{equation}
\begin{array}{lcr}
P_3 & = & P_l (n) \times \frac{1}{n \times \Delta P_l (n)} \\
P_3 & = & P_r (n) \times \frac{1}{n \times \Delta P_r (n)} ,\\
\end{array}
\label{eq3}
\end{equation}

where $\Delta P_l (n) = P_l (n) - P_1 (n)$, and $\Delta P_r (n) = P_1 (n) - P_r (n)$. Presence of the harmonics
satellites in the spectrum, apparently, reflects the presence of the beating of two periodic processes with period
$P_1$ and $P_3$. It is necessary to notice that both satellites are not in all pulsars, and in a case of presence
both, value of $P_3$ calculated from equation (3), is slightly more in the left companion (by $7-20\%$), but, as a
rule, both values coincide within a measuring error. It is necessary to note that in the paper  of Weltevrede et~al.~(\cite{W06}) also
there are the pulsars with the double value not only $P_2$, but also $P_3$. We will return to the our example.
Having calculated distance to the left and the right satellites at several harmonics of pulsar J0323+3944
(fig.2b), according to the equation (3), we calculated the period $P_3 = 8.4 \pm 1.3$. Weltevrede et~al.~(\cite{W06}) give the same
quantity $P_3 = 8.4 \pm 0.1$.

\begin{table}
\caption{Estimates of  the periods $P_2$ and $P_3$ from Fourier power spectra and according to  Weltevrede et~al.~(\cite{W06})}
\bigskip
\begin{tabular}{cccccc}
\hline
name      & $P_1$  &  $P_{2(LPA)}$  & $P_{2(W)}$  & $P_{3(LPA)}$ & $P_{3(W)}$    \\
      &  (s) &   (ms) &  (ms) &  &     \\
\hline
J0323+3944 &	3.03 &	112$\pm$12&	$152^{+42}_{-25}$ &		8.4$\pm$1.3 &		8.4$\pm$0.1\\[7pt]
J0528+2200 &	3.75 &	200$\pm$10&	$-208^{+20}_{-90}$ &	        4.9$\pm$0.5 &	       3.8$\pm$0.7\\[7pt]
	  &	    &	& $520^{+580}_{-105}$           &			&               3.7$\pm$0.4\\[7pt]
J0826+2637 &	0.53 &	150$\pm$25 &	 $80^{+60}_{-10}$ &		5.7$\pm$0.4 &		7$\pm$2\\[7pt]
J1136+1551 &	1.19&	400$\pm$70&	$430^{+400}_{-50}$&		 	&	        3$\pm1$\\[7pt]
	  &	    &	       &       $660^{+180}_{-300}$ &	                &                   \\[7pt]			
J1239+2453 &	1.38 &	60$\pm$3  & $61^{+4}_{-9}$	  &	2.7$\pm$0.1		&          2.7$\pm$0.1 \\[7pt]
	  &         &          &	$77^{+5}_{-11}$	  &			&                  \\[7pt]
J1813+4013 &	0.93 &	170$\pm$15&		  &			&                  \\[7pt]
J1907+4002 &	1.24 &	65$\pm$4  &		  &			&                  \\[7pt]
	  &         &	250$\pm$50&		  &			&                  \\[7pt]
J1921+2153 &	1.34 &	33$\pm$1  &	$13^{+1}_{-1}$    &		4.2$\pm$0.5&	4.4$\pm$0.1   \\[7pt]
	  &	    & 100$\pm$3   &	41$\pm$4	 &			&                  \\[7pt]
J2018+2839&	0.56&	& $-19^{+2}_{-12}$	  &	4.1$\pm$0.4	&	4$\pm$4       \\[7pt]
	  &	    &   93$\pm$8       & $-108^{+23}_{-44}$       &			& 	           \\[7pt]
J2022+2854&	0.34&	26$\pm$3  & $24^{+14}_{-2}$	  &			&    2.3$\pm$0.1      \\[7pt]
	  &	    &	       & $-52^{+5}_{-14}$         &			& 2.5$\pm$0.2         \\[7pt]
J2113+2754&	1.2 &	200$\pm$15& $470^{+60}_{-50}$	  &			& 4.4$\pm$0.1         \\[7pt]
J2157+4017&	1.53&	300$\pm$40& $470^{+370}_{-45}$	  &	4.5$\pm$0.7	&	3.1$\pm$0.8   \\[7pt]
J2305+3100&	1.58&	190$\pm$30& $66^{+14}_{-1}$	  &	2.2$\pm$0.4	&	2.1$\pm$0.1   \\[7pt]
J2317+2149&	1.44&	470$\pm$30&		  &			&                  \\
\hline
\end{tabular}
\end{table}

In table 1 the information on 14 pulsars from our sample is presented.
These pulsars are common with the list of the objects of Weltevrede et~al.~(\cite{W06}). In the first column of the table names of
the pulsars in J2000 are given. We give in the second column a pulsar period from the ATNF catalogue, in the
third column period $P_2$, and in the fifth column period $P_3$ of the pulsars, measured in our observations. For
comparison in the columns four and six the data from the paper Weltevrede et~al.~(\cite{W06}) are given. The errors of the measuring
$P_2$ are related to an accuracy of the definition of the number of a harmonic at which the middle of the first
maximum of the amplitude modulation is appeared. The error of its definition, as a rule, is $\pm 0.5$  numbers of a
harmonic of the basic spectrum. The error of  a measuring $P_3$ is related to the dispersion of the values $P_l (n)$
and $P_r (n)$, measured, not less than, for the three satellites at the several first harmonics. As a rule, the
number of the measures made from 6 to 20. The first measurements of $P_2$ and $P_3$ for 13 pulsars are presented
in tab.2. These pulsars have not included into the paper Weltevrede et~al.~(\cite{W06}). In the first, second and third columns the
same quantities, as in tab. 1 are given, and in the fourth one values $P_3$ are presented. $P_3$ given in tab.1, 2,
obtained in the main, as the mean value between $P_l (n)$ and $P_r (n)$.

The comparison of our estimates of the periods $P_2$ and $P_3$ with data of Weltevrede et~al.~(\cite{W06}) shows their good agreement.
Moreover, on our estimates of period $P_2$, for some pulsars, it is possible to remove of the not single-valued
in a direction of a subpulses motion in a pulse window. Our accuracy of  an evaluation of $P_2$ in most cases is
high, and for $P_3$ we see the comparable values.

\begin{table}
\caption{Estimates of $P_2$ and $P_3$ for the pulsars which have not included into the paper Weltevrede et~al.~(\cite{W06})}
\bigskip
\begin{tabular}{cccccc}
\hline
name      & $P_1$ (s) &  $P_{2(LPA)}$ (ms) &  $P_{3(LPA)}$   \\
\hline
J0611+3016&	1.41&	                 & 2.5$\pm$0,2 \\
J0613+3722&	0.29&	180$\pm$70	         &          \\
J0928+3037&	2.09&	280$\pm$20	         & 4.8$\pm$0,5 \\
J1532+2745&	1.12&	39$\pm$2	         & 4.6$\pm$0,5 \\
J1635+2418&	0.49&	140$\pm$30	         & 2.0$\pm$0,15\\
J1741+2758&	1.36&	$270^{+120}_{-60}$	 &          \\[7pt]
J1758+3030&	0.95&	105$\pm$15          &          \\	
J1912+2525&	0.62&	70$\pm$8	         &          \\
J2055+2209&	0.81&	80$\pm$10	         &          \\
J2139+2242&	1.08&	160$\pm$20	         &          \\
J2207+4057&	0.63&	210$\pm$40	         &          \\
J2227+3030&	0.84&	105$\pm$15	         &          \\
J2234+2114&	1.38&	230$\pm$20	         & 23.3$\pm$3,5\\
\hline
\end{tabular}
\end{table}

\section{Summary and discussion}

 The investigations of the integrated Fourier power spectra became possible, because there was data on four-year
monitoring of the large part of a sky. The volume of the data  has exceeded already 100 terabyte. The continuous
time of an accumulation for each point in the sky, after rejection of the data with the interferences, exceeds two
days. This rich data has given already a series of interesting results on the solar wind investigation
(Chashei et~al.\cite{Ch15}, Shishov et~al.~\cite{Sh16}), on the search for pulsars (Tyul'bashev et~al.~\cite{T16}, Tyul'bashev et~al.~\cite{T17a}) and on the search for the prompt radio transients
(Tyul'bashev \& Tyul'bashev~\cite{TT17}). The integrated power spectra  reveal also two new, unexpected features which we interpret
as a display of  the periodical processes in the field of radiation of the pulsar pulses.

If our interpretation is
true for all 14 pulsars in tab. 1 we have obtained an estimate of $P_2$. For 7 pulsars it well agree with the
data given by Weltevrede et~al.~(\cite{W06}), and for 4 sources there is a discrepancy in the measurements. Thus for three pulsars
from these four (J1921+2153, J2157+4070 and J2305+3103) values of $P_3$ are coincide with Weltevrede et~al.~(\cite{W06}). Only for
J2113+2754 we do not see the satellites at the first harmonics. For two pulsars we cannot exactly determinate
the first "hump" on a spectrum. Therefore for pulsars J1907+4002 and J1921+2153 we give two values of this quantity
(Tab.1). For 3 pulsars (J1813+4013, J1907+4002 and J2317+2149) we made for the first time the estimate of $P_2$,
because in the paper Weltevrede et~al.~(\cite{W06})  there are no this data.  Such
situation can indicate that or the additional investigations of these pulsars are necessary, which do not have
agreement, or the nature of the modulations which we observe  is related not only to a period $P_2$.
The values of $P_3$ has confirmed for 8 pulsars from 11 of list of Weltevrede et~al.~(\cite{W06}), and in three remained sources we
do not see already the significant harmonics of the satellites in the integrated power spectrum (tab.1).
The data for 13 pulsars, given in tab. 2 are new and they give estimates of  $P_2$ for 12 sources and
estimates of $P_3$ for 5 pulsars, including pulsar J0928+3037 recently discovered by us (Tyul'bashev et~al.~\cite{T16}). The values
$P_2$ are in a range 26-470 ms and the values $P_3$ lie in a range from 2 to 23.3 $P_1$ (Tab.1-2). All new
estimates of both $P_2$ and $P_3$ need in the confirmation by other methods.

Thus, as a result of the analysis of the integrated power spectra obtained by a summation of 500-600 daily
spectra, their some new features are discovered. It is shown that two of them, namely a modulation with the
periods from 3 to 40 first harmonics  is reflected to a period of the second class ($P_2$), and a presence of
the harmonic satellites is testified of a period of the third class ($P_3$). We find the empirical relations
(1-3) for the calculation of both periods. As a result of the analysis, a measuring of these periods and a
comparison with the data of (Weltevrede et~al.~\cite{W06})  the value $P_2$ is estimated  for 26
pulsars, and  for 15 sources it is made for the first time. The value $P_3$ is estimated for 13 pulsars, from
them for 5 sources they are given for the first time.

\section{Acknowledgements}
The authors are grateful for support of the grant of the Russian Foundation for Basic  Research  16-02-00954 and
for an active participation of V.S.Tyulbashev, V.V.Oreshko and S.V.Logvinenko. Authors express gratitude
L.B.Potapova for the help with the paper design and I.F.Malov for the help with the translation.


\begin{thebibliography}{99}
\bibitem[2013]{B13}	
Barr E.D. et al., 2013, MNRAS, 435, 2234
\bibitem[2015]{Ch15}	
Chashei I.V., Shishov V.I., Tyul'bashev S.A., Subaev I.A., Oreshko V.V., Logvinenko S.V., 2015, \solphys, 290, 2577
\bibitem{www}	
Catalogue ATNF - http://www.atnf.siro.au
\bibitem[1999]{DR99}	
Deshpande A.A. \& Rankin J.M., 1999, \apj, 524, 1008
\bibitem[1968]{DC68}	
Drake F.D. \& Craft H.D., 1968, \nat, 220, 231
\bibitem[2000]{M00}
Malofeev V.M., Malov O.I., Schegoleva N.V., 2000, Astron. Rep., 44, 436
\bibitem[1975]{OS75}	
Sieber W. \& Oster L., 1975, \aap, 38, 325
\bibitem[2003]{RSD03}	
Rankin J.M., Suleymanova S.A., Deshpande A.A., 2003, MNRAS, 340, 1076
\bibitem[1968]{R68}	
Ruderman M.A., 1968, \nat, 218, 1128
\bibitem[1975]{RS75}	
Ruderman M.A. \& Sutherland P.G., 1975, \apj, 196, 51
\bibitem[2016]{Sh16}		
Shishov V.I., et al., 2016, Astron. Rep., 60, 1067
\bibitem[2016]{T16}	
Tyul'bashev S.A., Tyul'bashev V.S., Oreshko V.V., Logvinenko S.V., 2016, Astron. Rep., 60, 220
\bibitem[2017]{T17a}	
Tyul'bashev S.A. et al., 2017, Astron. Rep., 61, 848
\bibitem[2017a]{TT17}	
Tyul'bashev S.A., Tyul'bashev V.S., 2017a, ATsir., 1636, 1
\bibitem[1970]{V70}	
Vitkevich V.V. \& Shitov Yu.P., 1970, \nat, 225, 248
\bibitem[2006]{W06}	
Weltevrede P., Edvards R.T., Stappers B.W., 2006, \aap, 445, 243
\bibitem[2003]{W03}	
Wright G.A.E., 2003, MNRAS, 344, 1041
\end{thebibliography}
\end{document}